# Polarized Gamma-ray Emission from the Galactic Black Hole Cygnus X-1

P. Laurent[1,*], J. Rodriguez[2], J. Wilms[3], M. Cadolle Bel[4], K. Pottschmidt[5,6], V. Grinberg[3]

[1]*Astroparticules et Cosmologie (APC), CEA/IRFU, 10, rue Alice Domon et Léonie Duquet, 75205, Paris Cedex 13, France;* [2]*Laboratoire AIM, CEA/IRFU, Université Paris Diderot, CNRS/INSU, CEA Saclay, DSM/IRFU/SAp, 91191 Gif sur Yvette, France;* [3]*Dr. Karl Remeis-Sternwarte and Erlangen Centre for Astroparticle Physics, Friedrich-Alexander-Universität Erlangen-Nürnberg, Sternwartstr. 7, 96049 Bamberg, Germany;* [4]*INTEGRAL Science Operations Centre, Science Operations Department, ESAC, P.O. Box 78, E-28691 Villanuevade la Cañada, Madrid, Spain;*[5]*Center for Research and Exploration in Space Science & Technology (CRESST) and National Aeronautics and Space Administration Goddard Space Flight Center, Astrophysics Science Division, Code 661, Greenbelt, MD 20771, USA;* [6]*Center for Space Science & Technology, University of  Maryland Baltimore County, 1000 Hilltop Circle, Baltimore, MD 21250,  USA*

[*]To whom correspondence should be addressed. E-mail: plaurent@cea.fr

**Because of their inherently high flux allowing the detection of clear signals, black hole X-ray binaries are interesting candidates for polarization studies, even if no polarization signals have been observed from them before. Such measurements would provide further detailed insight into these sources' emission mechanisms. We measured the polarization of the gamma-ray emission from the black hole binary system Cygnus X-1 with the INTEGRAL/IBIS telescope. Spectral modeling of the data reveals two emission mechanisms: The 250-400 keV data are consistent with emission dominated by Compton scattering on thermal electrons and are weakly**

**polarized. The second spectral component seen in the 400keV-2MeV band is by contrast strongly polarized, revealing that the MeV emission is probably related to the jet first detected in the radio band.**

Cygnus X-1 is probably the best known black hole (BH) X-ray binary in our Galaxy. It has been widely observed with many telescopes over the whole electromagnetic band (*1-10*). The BH is located around 2.1 kpc away from Earth (*11*), and forms a binary system with a high mass blue O star (*12*). It radiates mainly in the X-ray and soft gamma-ray domains; the X-ray luminosity is thought to be produced by accretion of the companion's matter onto the BH (*1,2*). The well-studied X-ray spectrum is a combination of a thermal spectrum with temperature around 130 eV (*13*) and a cutoff power law spectrum, due to the Compton-scattering of the disk photons off high temperature thermal electrons located in a corona close to the BH (*2*). Recently, an additional spectral component of unknown origin was observed (*10*) by the spectrometer on INTEGRAL (SPI) telescope (*14*). Finally, a compact radio jet is ejected from the vicinity of the BH, with a kinetic power similar to the source's bolometric X-ray luminosity (*3,4*).

The IBIS telescope (*15*) onboard the INTEGRAL satellite (*16*), can be used as a Compton polarimeter (*17-21*). Spectral measurements of Cygnus X-1 (Fig. 1) reveal two high energy components: a cutoff power law component between 20 and 400 keV, reminiscent of a Compton-scattering induced spectrum, already observed by many satellites (*2,6,7,9,10*), and a power law spectrum at higher energies of up to 2 MeV, already observed (*10*) by the SPI telescope on board INTEGRAL. These two components are signatures of two different high energy emission processes from the source, whose locations have not been previously constrained.

We measured the polarization signal between 250 and 400 keV (Figure 2). As expected from a zone where Compton scattering on thermal electrons dominates (*22*), the emission in this band is weakly polarized with an upper limit of 20% for the polarization fraction Pf.

In contrast, the signal from the 400-2000keV band, in which the hard tail dominates, is highly polarized (Pf = 67 ± 30 %, see Figure 3). This result is no longer consistent with Compton scattering on thermal electrons (*22*), and such a high polarization fraction is probably the signature of synchrotron or inverse Compton emission from the jet already observed in the radio band (*23*). Unfortunately, current knowledge of the jet at radio wavelengths does not allow discriminating between the two processes.

In order to have such a clear polarimetric signal, the magnetic field has to be coherent over a large fraction of the emission site (*5*). Such a coherent magnetic field structure may indicate a jet origin for the gamma-rays above 400 keV (*24*). In addition, because the gamma-rays emitted in BH X-ray binaries are generally thought to be emitted close to the BH horizon (*7,25*), and because the synchrotron photons we observed in the hard tail are too energetic to be effectively self-Comptonized, these observations might be evidence that the jet structure is formed in the BH vicinity, possibly in the Compton corona itself. Another possibility is that the gamma-rays are produced in the initial acceleration region in the jet, as observed at higher energies by Fermi/LAT from the microquasar Cygnus X-3 (*26*).

The spectrum observed above 400 keV is consistent with a power law of photon index 1.6 ± 0.2. This means that this spectrum, if due to synchrotron or inverse Compton emission, is caused by electrons whose energy distribution is also a power law with an index p of 2.2 ± 0.4 (*27*), consistent with the canonical value for shock-accelerated particles p = 2. Synchrotron radiation at MeV energies implies also that the electron energy, for a magnetic field of 10 mG, which is reasonable for this kind of system (*28*), would be around a few TeV (*27,29*). Inverse Compton scattering of photons off these high energy TeV electrons, whose lifetime due to synchrotron energy loss is around one month (*27*), could also be the origin of the TeV photons detected from Cygnus X-1 with the MAGIC experiment (*30*) and possibly also the gamma-rays claimed by AGILE (*31*).

The position angle (PA) of the electric vector, which gives the direction of the electric field lines projected onto the sky, is 140 ± 15°. This is at least 100° away from the compact radio jet, which is observed at a PA of 21-24° (*32*). Such deviations between the electric field vector and jet direction are also found in other jet sources, such as Active Galactic Nuclei (*33*) or the galactic source SS433 (*34*).

21. In brief, a Compton polarimeter utilizes the polarization dependency of the differential cross section for Compton scattering, where linearly polarized photons scatter preferentially perpendicularly to the incident polarization vector. By examining the scatter angle azimuthal distribution of the detected photons a sinusoidal signal is obtained from which the polarization angle PA and the polarization fraction Pf with respect to a 100 % polarized source can be derived. Gamma-ray polarization measurements are particularly difficult, the main difficulty being the exclusion of systematic/detector effects in the azimuthal Compton events distribution. To exclude these systematic effects, we followed the process detailed in (*19*). We only considered events that interacted once in the upper CdTe crystal layer, ISGRI (sensitive in the 15-1000keV band), and once in the lower CsI layer, PICsIT (sensitive in the 200keV-10MeV band), and whose reconstructed energy was in the 250 keV-2000 keV energy range. These events were automatically selected on board through a time coincidence algorithm. The maximal allowed time window was set to 3.8 µs during our observations, which span between 2003 and 2009, for a total exposure of more than 5 million seconds, which is around 58 days.

35. ISGRI has been realized and maintained in flight by CEA-Saclay/Irfu with the support of CNES. Based on observations with INTEGRAL, an ESA project with instruments and science data centre funded by ESA member states (especially the PI countries: Denmark, France, Germany, Italy, Switzerland, Spain), Czech Republic and Poland, and with the participation of Russia and the USA. We acknowledge partial funding from the European Commission under contract ITN 215212 "Black Hole Universe" and from the Bundesministerium für Wirtschaft und Technologie under Deutsches Zentrum für Luft- und Raumfahrt grant 50 OR 1007. K. Pottschmidt acknowledges support by NASA's

INTEGRAL Guest Observer grants NNX08AE84G, NNX08AY24G, and NX09AT28G.

We thank S. Corbel for useful comments.

**Fig. 1.** Cygnus X-1 energy spectrum as measured by the Integral/IBIS telescope, and obtained with the standard IBIS spectral analysis pipeline. Two components are clearly seen: a "Comptonisation" spectrum caused by photons upscattered by Compton scattering off thermally distributed electrons in a hot plasma (dashed line), and an higher energy component (dash dot line) whose origin is not known.

**Fig. 2.** Cygnus X-1 polarization signal measured in two adjacent energy bands. This distribution gives the source count rate by azimuthal angle of the Compton scattering. In the 250-400 keV energy band (panel a), the signal is consistent with a flat signal indicating that the observed gamma-rays are weakly or even not polarized. In the 400-2000 keV energy band (panel b), the signal is now highly modulated, indicating that the observed gamma-rays are highly polarized.

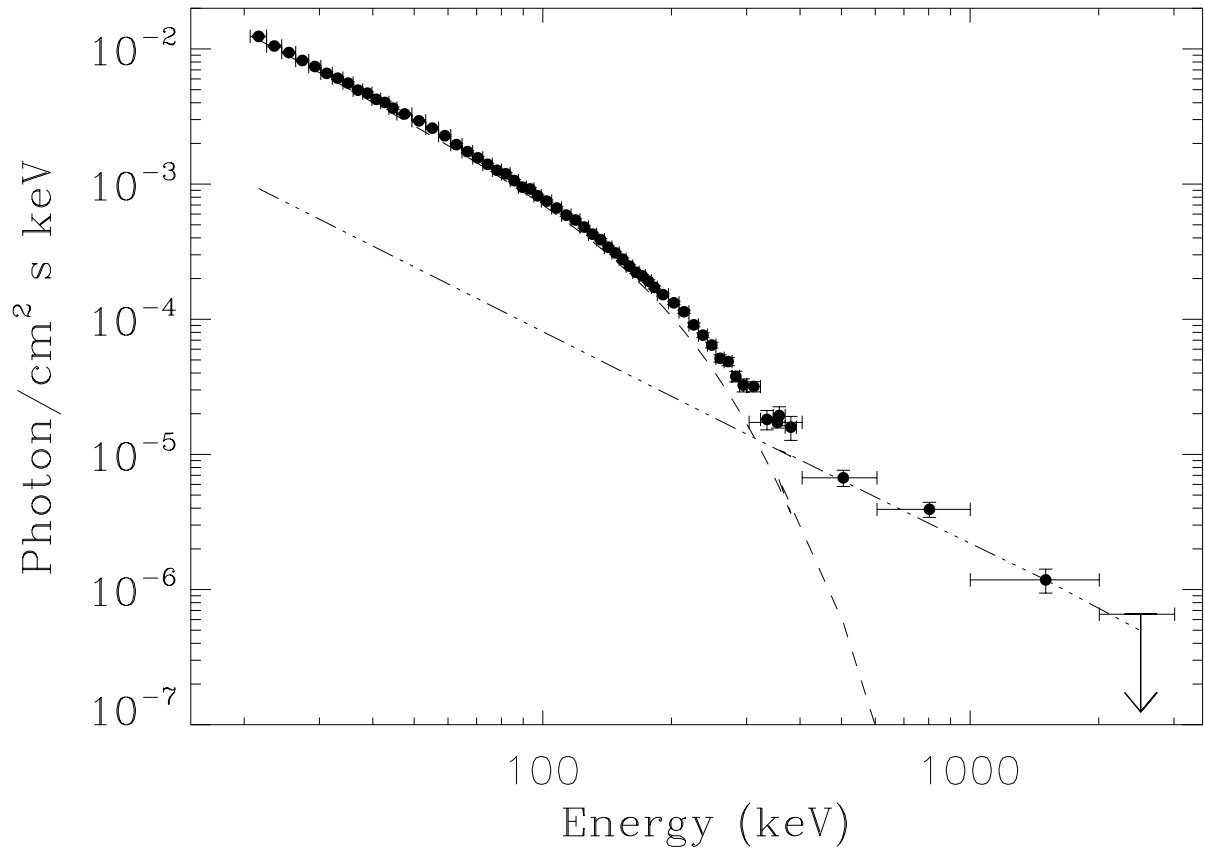

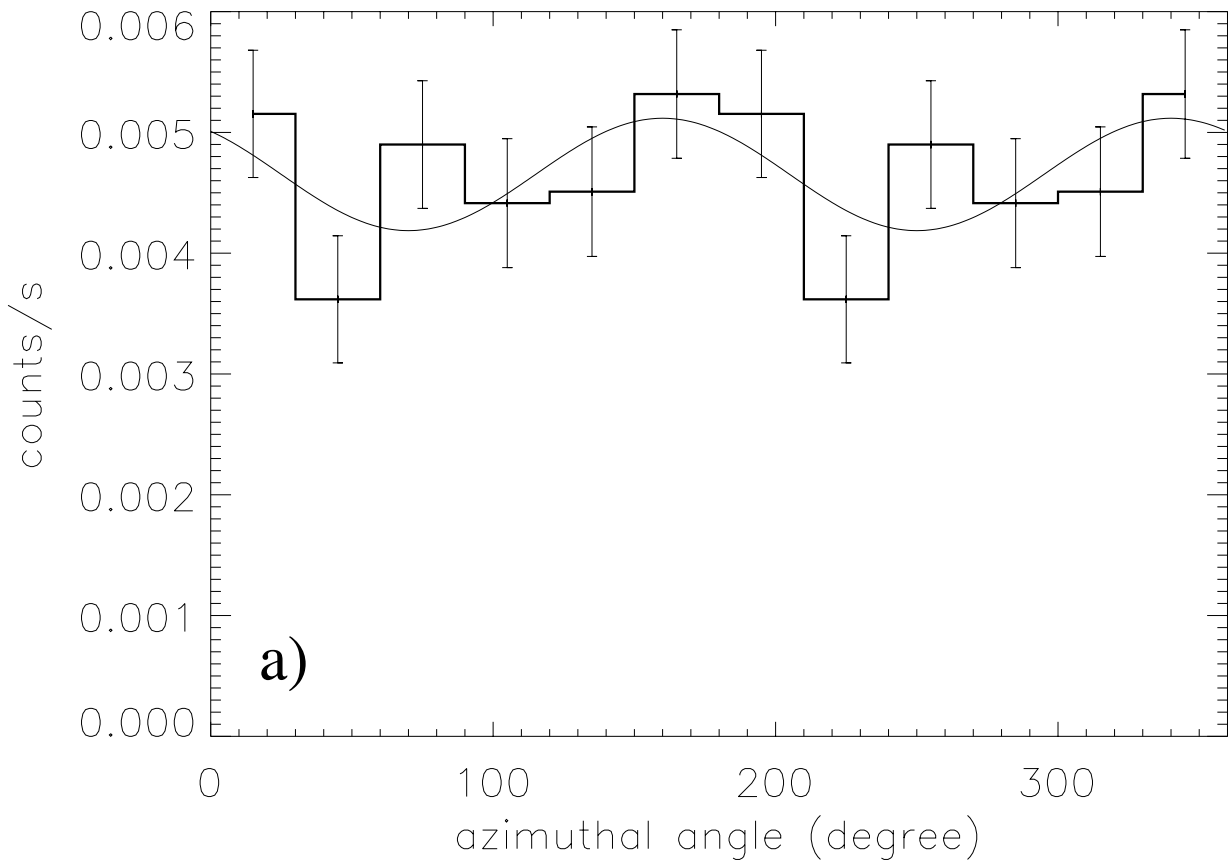
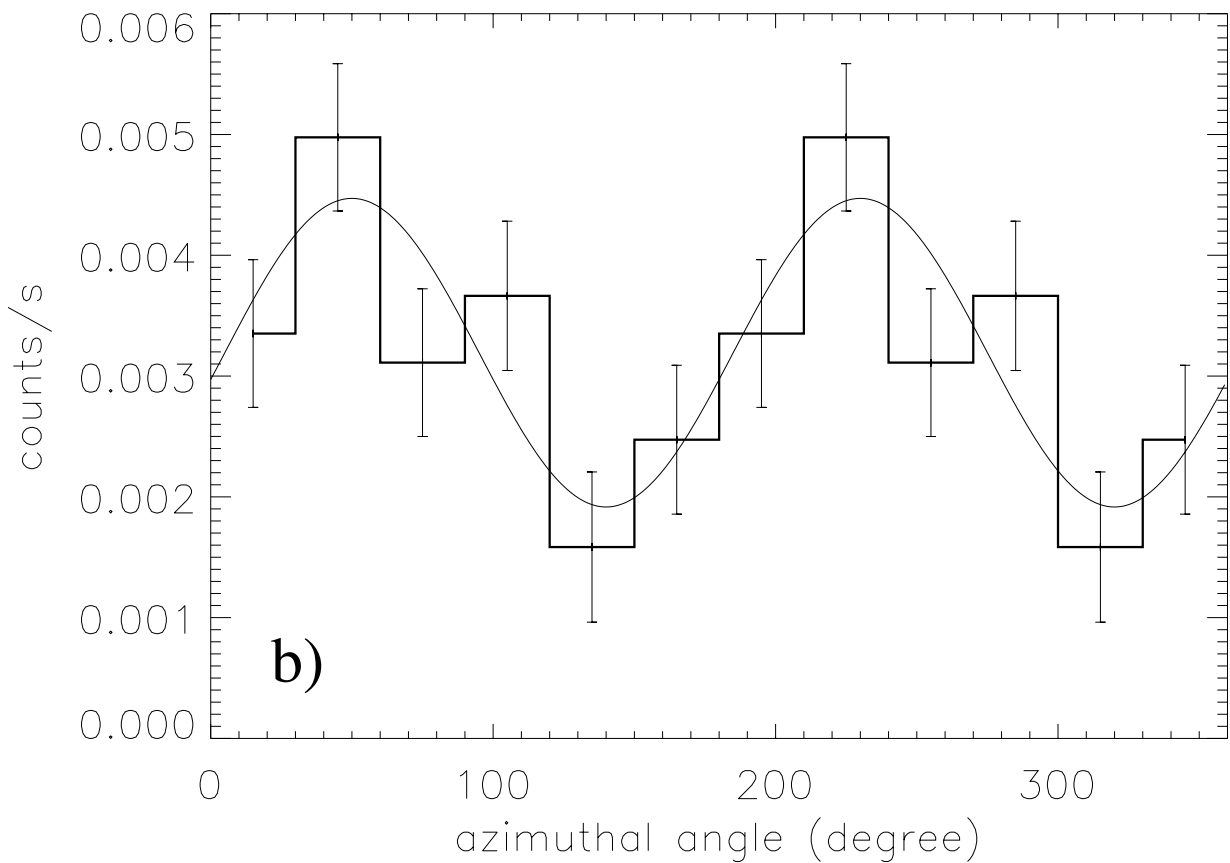